\documentclass[aps,pra,reprint,nobibnotes,nofootinbib,floatfix]{revtex4-2}
\usepackage[T1]{fontenc}
\usepackage{lmodern}
\usepackage{textcomp}
\usepackage{amsmath,booktabs}
\usepackage{tikz}
\usetikzlibrary{arrows.meta}
\usepackage[hidelinks]{hyperref}
\hypersetup{
  pdftitle={Sub-2 \textmu s round trip between an FPGA and a decoding host for quantum error correction},
  pdfauthor={X. Chin, L. Deng, H. Liu, R. van Staden, S. Zhang},
  pdfsubject={Design and latency budget of a 100 GbE RoCEv2 remote-procedure-call link between a control FPGA and a decoding host}}

\usepackage[stretch=30,shrink=30]{microtype}
\usepackage{etoolbox}
\makeatletter
\@booleantrue\raggedcolumn@sw
\def\section{\@startsection{section}{1}{\z@}{16pt \@plus 2pt \@minus 1pt}{7pt}{\normalfont\small\bfseries\centering}}
\def\subsection{\@startsection{subsection}{2}{\z@}{12pt \@plus 2pt \@minus 1pt}{5pt}{\normalfont\small\bfseries\centering}}
\def\balance@even#1{%
 \setbox\z@\copy\@ne\setbox\tw@\vsplit\z@ to\dimexpr\dimen@+#1\relax
 \remove@depth\z@\remove@depth\tw@
 \@tempdima\ht\tw@\advance\@tempdima-\ht\z@
 \@ifdim{\@tempdima<\z@}{\@tempdima-\@tempdima}{}%
 \@tempdimb\dimen@ii
 \@ifdim{\@tempdimb<\z@}{\@tempdimb-\@tempdimb}{}%
 \@ifdim{\@tempdima<\@tempdimb}{}{%
  \setbox\z@\copy\@ne\setbox\tw@\vsplit\z@ to\dimen@
  \remove@depth\z@\remove@depth\tw@
 }%
}
\def\balance@pick{%
 \@ifdim{\dimen@ii<-\p@}{\balance@even{2\p@}}{%
  \@ifdim{\dimen@ii>\p@}{\balance@even{-2\p@}}{}}%
}
\patchcmd\balance@two{\@ifdim{\ht\z@=\z@}}{\balance@pick\@ifdim{\ht\z@=\z@}}{}{}
\def\bib@room{%
 \par
 \@ifdim{\pagegoal<\maxdimen}{%
  \@ifdim{\dimexpr\pagegoal-\pagetotal\relax<100\p@}{\newpage}{}%
 }{}%
}
\AtBeginDocument{\patchcmd\bibsection{\onecolumngrid@push}{\onecolumngrid@push\bib@room}{}{}}
\long\def\@makecaption#1#2{%
  \par
  \vskip\abovecaptionskip
  \begingroup
   \footnotesize\rmfamily
   \sbox\@tempboxa{%
    \let\\\heading@cr
    \@make@capt@title{#1}{#2}%
   }%
   \@ifdim{\wd\@tempboxa >\hsize}{%
    \begingroup
     \samepage
     \flushing
     \let\footnote\@footnotemark@gobble
     \@make@capt@title{#1}{#2}\par
    \endgroup
   }{%
     \global \@minipagefalse
     \hb@xt@\hsize{\hfil\unhbox\@tempboxa\hfil}%
   }%
  \endgroup
  \vskip\belowcaptionskip
}%
\AtBeginDocument{%
  \setlength{\textfloatsep}{24pt plus 2pt minus 2pt}%
  \setlength{\dbltextfloatsep}{24pt plus 2pt minus 2pt}%
  \raggedbottom
  \hyphenpenalty=10000 \exhyphenpenalty=10000
  \tolerance=2000 \emergencystretch=2em
  \clubpenalty=10000 \@clubpenalty=10000
  \widowpenalty=10000 \displaywidowpenalty=10000}
\makeatother

\newcommand{\us}{\textmu s}
\newcommand{\rt}[1]{\mbox{#1\,\textmu s}}
\newcommand{\ns}[1]{\mbox{#1\,ns}}
\newcommand{\enq}{\mbox{\texttt{enqueue\_syndromes}}}
\newcommand{\getc}{\mbox{\texttt{get\_corrections}}}
\newcommand{\contact}[1]{\href{mailto:#1@taiyiliangsheng.com}{#1@taiyiliangsheng.com}}

\newlength{\figu}
\definecolor{taiyiblue}{HTML}{0049E6}
\definecolor{taiyinavy}{HTML}{000086}
\definecolor{taiyipale}{HTML}{DDE5F0}
\definecolor{taiyiline}{HTML}{8A94A6}
\definecolor{taiyitext}{HTML}{4A5568}
\colorlet{figline}{taiyiline}        
\colorlet{figquiet}{taiyitext}       
\colorlet{figdark}{taiyinavy}        
\colorlet{figmid}{taiyiblue}
\colorlet{figsky}{taiyiblue!35!white}
\colorlet{figwarm}{taiyipale}
\colorlet{figaccent}{taiyiblue}      
\colorlet{figpale}{taiyipale}        
\tikzset{
  fig/.style={line cap=round, line join=round,
    >={Triangle[length=4.2pt, width=4.2pt]},
    every node/.style={inner sep=0pt, outer sep=0pt, anchor=base,
      font=\sffamily\scriptsize, text=figquiet}},
  fighead/.style={font=\sffamily\footnotesize\bfseries, text=black},
  figink/.style={text=black},
  figbox/.style={draw=figline, line width=0.8pt, rounded corners=4pt},
  figgroup/.style={draw=figline, line width=0.6pt, rounded corners=4pt},
  figmain/.style={draw=figaccent, fill=figpale, line width=1.2pt, rounded corners=4pt},
  figone/.style={draw=figline, line width=0.8pt, ->},
  figtwo/.style={draw=figline, line width=0.8pt, <->},
  figbr/.style={draw=figline, line width=0.7pt, line cap=butt, line join=miter},
  figinst/.style={anchor=mid, text=black, fill=white, inner xsep=3pt, inner ysep=0.5pt},
}

\begin{document}
\title{Sub-2\texorpdfstring{\,}{ }\textmu s round trip between an FPGA and a decoding host for quantum error correction}
\author{X.~Chin}
\thanks{\raggedright Authors are listed in alphabetical order.}
\affiliation{Taiyi Quantum Science \& Technology Co.\ Ltd.}

\author{L.~Deng}
\affiliation{Taiyi Quantum Science \& Technology Co.\ Ltd.}

\author{H.~Liu}
\thanks{\raggedright Corresponding author: \contact{hbliu}.}
\affiliation{Taiyi Quantum Science \& Technology Co.\ Ltd.}

\author{R.~van~Staden}
\thanks{\raggedright Corresponding author: \contact{rvanstaden}.}
\affiliation{Taiyi Quantum Science \& Technology Co.\ Ltd.}

\author{S.~Zhang}
\affiliation{Taiyi Quantum Science \& Technology Co.\ Ltd.}
\date{24 November 2026}
\begin{abstract}
Real-time quantum error correction places a classical round trip inside the control loop: syndrome data leave the control electronics, a decoder responds, and the decision must return before the next dependent gate. NVIDIA’s NVQLink demonstrated that a quantum controller could exchange a short message with a GPU over Ethernet in a few microseconds. Building on that work, we customize the controller datapath for short syndrome messages and measure how much time can be saved on a round trip. A control field-programmable gate array (FPGA) issues 64-byte CUDA-Q remote procedure calls to an NVIDIA DGX Spark by remote direct memory access (RDMA) over Converged Ethernet version 2 (RoCEv2), on a 100\,Gb/s link. On the Spark side, a single polling thread calls the relay, invokes a CUDA-Q decoding session and replies with one RDMA SEND. Every round trip is timed in the FPGA. A fast path of custom transmit and receive logic on the clock of the 100\,Gb/s Ethernet media access controller of the FPGA (CMAC) reduces the latency contribution of the FPGA logic from \ns{238} to \ns{12}. Through the decoding session, the median latencies of nine bursts of 200 calls on the fast path lie between \rt{1.87} and \rt{1.95}. In a like-for-like test, with the same RDMA operating as the HSB reference path, the fast path median latency of the round trip is reduced from \rt{2.23} to \rt{1.99}. Independent measurements attribute \ns{285} of that latency to the CMAC and transceivers, and \rt{1.44} to the port-to-port turnaround of the Spark itself, i.e., about 85\% of the round trip is spent beyond the CMAC. The Spark, rather than the protocol logic of the FPGA or the host software, now limits the link.
\end{abstract}
\maketitle

\section{Introduction}
Fault-tolerant quantum computing requires that the syndrome measurements of quantum error correction (QEC) be decoded in parallel to the computational operations. A decoder that cannot keep pace with the syndrome stream accumulates a backlog that grows without bound~\cite{terhal}, and the response time of a decoder sets the pace of every logical operation that waits for its result~\cite{skoric}. Gidney and Eker\aa{} budget \rt{10} from triggering a logical measurement to choosing the basis of the next one in their resource estimate for superconducting qubits~\cite{gidney}, and Kurman et al.\ derive that the closed loop between controller and decoder must stay within tens of microseconds~\cite{kurman}. On superconducting hardware a syndrome round already takes about a microsecond~\cite{willow}, and in a demonstrated real-time experiment communication and control took about a third of the time from final readout to conditional feedback~\cite{caune}.

\begin{figure*}[t]
\centering
\setlength{\figu}{\dimexpr\textwidth/760\relax}%
\begin{tikzpicture}[fig, x=\figu, y=-\figu]
  \useasboundingbox (0,22) rectangle (760,274);
  \draw[draw=figline, line width=0.8pt, line cap=butt, dash pattern=on 2.6pt off 2.6pt]
    (472,92) -- (472,40) -- (198,40) -- (198,92);
  \node at (335,32) {Control plane: BOOTP and register access over UDP, through HSB};
  \draw[figgroup] (4,56) rectangle (356,250);
  \node[fighead, anchor=base west] at (18,77) {Control FPGA (VCU118)};
  \draw[figgroup] (420,56) rectangle (756,250);
  \node[fighead, anchor=base east] at (742,77) {DGX Spark};
  \draw[figgroup, fill=figwarm] (526,150) rectangle (748,243);
  \node[fighead, anchor=base east] at (742,165) {Host Software};
  \draw[figtwo] (112,125) -- (136,125);
  \draw[figtwo] (260,125) -- (284,125);
  \draw[figtwo] (112,203) -- (136,203);
  \draw[figtwo] (260,203) -- (284,203);
  \draw[figone] (342,140) -- (434,140);
  \draw[figone] (434,196) -- (342,196);
  \draw[figone] (510,116) -- (532,116);
  \draw[figone] (576,140) -- (576,174);
  \draw[figtwo] (614,205) -- (660,205);
  \draw[figone] (532,222) -- (510,222);
  \draw[figbox] (18,92) rectangle (112,236);
  \node[fighead] at (65,115) {RPC Client};
  \node[figink] at (65,144) {timer start};
  \node at (65,157) {request issued};
  \node[figink] at (65,190) {timer stop};
  \node at (65,203) {reply accepted};
  \draw[figbox] (136,92) rectangle (260,158);
  \node[fighead] at (198,109) {HSB};
  \node at (198,124) {250\,MHz};
  \node at (198,137) {two FIFOs};
  \node[figink] at (198,150) {adds 238\,ns};
  \draw[figmain] (136,170) rectangle (260,236);
  \node[fighead] at (198,187) {Fast Path};
  \node at (198,202) {CMAC clock, 322\,MHz};
  \node at (198,215) {no FIFO};
  \node[figink] at (198,228) {adds 12\,ns};
  \draw[figbox] (284,92) rectangle (342,236);
  \node[fighead] at (315,161) {CMAC};
  \node at (315,177) {100\,Gb/s};
  \fill[figaccent] (284,157) -- (289,164) -- (284,171) -- (279,164) -- cycle;
  \node[figink] at (388,132) {WRITE};
  \node[figink] at (388,211) {SEND};
  \draw[figbox] (434,92) rectangle (510,236);
  \node[fighead, font=\sffamily\scriptsize\bfseries] at (472,161) {ConnectX-7};
  \node[fighead] at (472,177) {Adapter};
  \draw[figbox] (532,92) rectangle (626,140);
  \node[fighead] at (579,112) {Ring};
  \node at (579,127) {Spark memory};
  \draw[figbox, fill=white] (532,174) rectangle (614,236);
  \node[fighead] at (573,203) {Relay};
  \node at (573,218) {one thread};
  \draw[figbox, fill=white] (660,174) rectangle (742,236);
  \node[fighead] at (701,194) {Decoding};
  \node[fighead] at (701,207) {Session};
  \node at (701,223) {CUDA-Q QEC};
  \node[figink] at (637,198) {C ABI};
  \fill[figaccent] (18,257) -- (23,264) -- (18,271) -- (13,264) -- cycle;
  \node[anchor=base west] at (30,267.5) {timestamps at the CMAC boundary};
\end{tikzpicture}
\caption{System architecture. A request crosses the direct 100\,Gb/s Ethernet link as one RDMA WRITE into a ring in the memory of the Spark, and the reply returns as one RDMA SEND. The relay calls the decoding session through a C ABI. The shaded group on the right is the host software; everything else there is the Spark itself. The labels on HSB and the fast path give the latency that each adds to a round trip. HSB carries the control plane (dashed) in both cases.}
\label{fig:system}
\end{figure*}
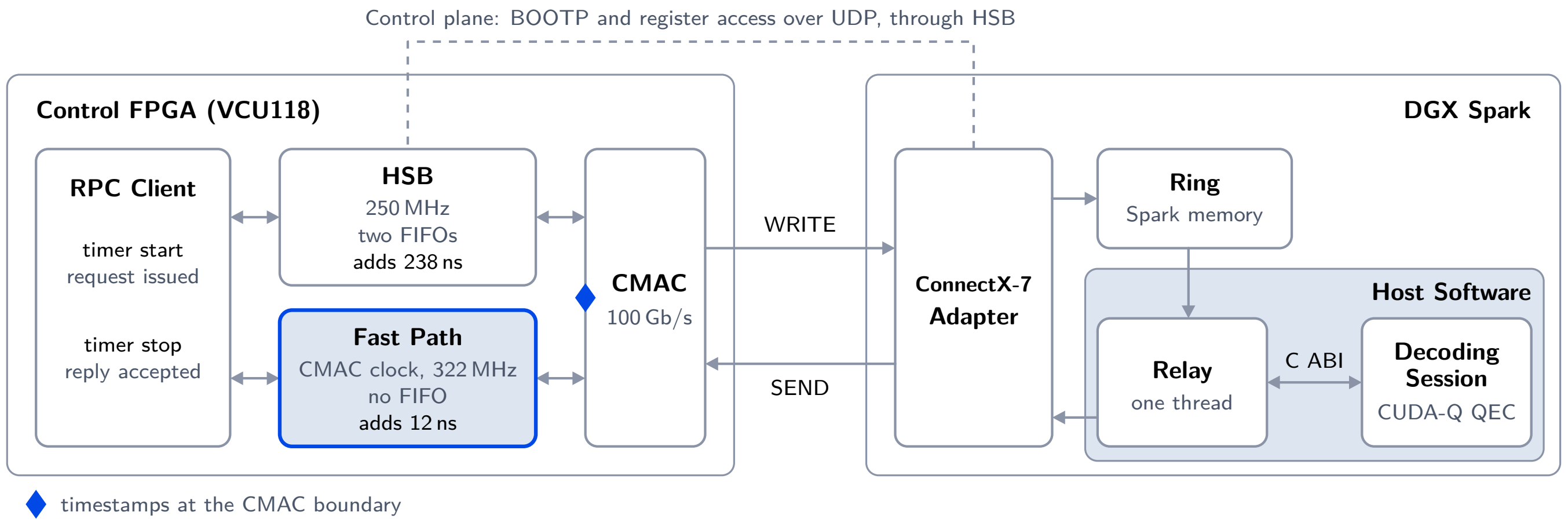

Decoders built into the control field-programmable gate array (FPGA) pay no transfer cost~\cite{barber,liyanage,wu,yang}, but is limited by the FPGA resource. Decoders on central processing units (CPUs) and graphics processing units (GPUs) are easier to change, are capable of dealing with larger models, and can be shared among controllers~\cite{nvq,backline,thq,qdpu,wang}. The data involved are small, e.g. a few bytes of ancilla qubits status per round for a distance-seven surface-code patch~\cite{wang}, and networking the controllers would allow FPGA boards in different chassis send independent blocks to shared the CPU/GPU servers as measurements arrive~\cite{nvq}. The price is that every decision received from the decoder now includes a network round trip.

How much of the reaction budget does that round trip consume? An exchange of less than \rt{2} leaves more than \rt{8} of a \rt{10} budget for readout, decoding and control. Published controller links report this cost as an end-to-end figure, typically for an echo or a forwarding test~\cite{nvq,backline,thq,qdpu}. To our best knowledge, none of the literature further disassemble the path into FPGA logic, Ethernet media access controller (MAC), adapter and software, and measure or discuss their respective contribution to the total latency. However, in order to make improvements on microsecond-scale systems argues such a latency budget is an essential starting point.~\cite{barroso}.

This paper describes a link between a control FPGA and a networked decoding machine and accounts for its latency. Requests are remote procedure calls (RPCs), carried by remote direct memory access (RDMA) in its Ethernet form, RDMA over Converged Ethernet version 2 (RoCEv2). Six terms are used throughout. The \emph{Spark} is the NVIDIA DGX Spark that answers the requests: the machine itself, with its network adapter, Peripheral Component Interconnect Express link, memory and processor. The \emph{host software} is what we run on the Spark to serve the requests. Its data path is the \emph{relay}, a single polling thread that takes each request from memory, calls (relays) the decoding session through a C application binary interface (ABI) and posts the reply. The FPGA sends and receives through its integrated 100\,Gb/s Ethernet MAC (\emph{CMAC})~\cite{cmac}. The \emph{fast path} is our transmit and receive logic on the clock of the CMAC. NVIDIA's \emph{Holoscan Sensor Bridge} (HSB)~\cite{hsb} is the FPGA intellectual-property (IP) core through which the same requests pass for comparison. The paper makes three contributions.

\emph{Design.} A request is a single RDMA WRITE into the memory of the Spark, and the reply is a single RDMA SEND. On the FPGA, the fast path builds each frame from a header template and derives its integrity check from the linearity of the cyclic redundancy check (CRC), so that no part of a frame is buffered (Sec.~\ref{sec:design}).

\emph{Measurement.} Every round trip is timed in the FPGA. Through a CUDA-Q decoding session the burst medians on the fast path lie between 1.87 and \rt{1.95}. In a like-for-like test against HSB, with the same RDMA operation, the fast path lowers the median from \rt{2.23} to \rt{1.99} (Sec.~\ref{sec:results}).

\emph{Budget.} Three independent measurements give \ns{12} of FPGA logic, \ns{285} in the CMAC and transceivers and \rt{1.44} for the port-to-port turnaround of the Spark; the small remainder holds the host software, which a fourth instrument times directly (Sec.~\ref{sec:budget}).

Two limits of scope should be stated at the outset. First, the decoding session runs an identity decoder, which returns its input, so the times reported are those of the link and of the session interface, not of decoding. Second, the responder is a CPU thread; the system has no path into GPU memory.

\section{System design}\label{sec:design}
\subsection{Architecture}
Figure~\ref{fig:system} shows the system. An AMD VCU118 board stands in for the control electronics. An RPC client in its logic issues requests, checks every reply and times each round trip. Requests leave through the CMAC and a direct 100\,Gb/s Ethernet (100\,GbE) cable to one port of the ConnectX-7 adapter of the Spark, a desktop system built on an Arm system-on-chip.

Each request is one RDMA WRITE into a small ring of registered memory on the Spark, one slot per request. The relay watches that memory, calls the handler that the request names, and answers with one RDMA SEND, which the FPGA receives. Neither the kernel nor an interrupt takes part in the exchange.

Two routes lead from the RPC client to the CMAC, and one design contains both. One is HSB, the FPGA IP on which NVIDIA's NVQLink architecture builds~\cite{nvq}. It forms the RoCEv2 frame at 250\,MHz, and transmit and receive first-in, first-out buffers (FIFOs) bridge that clock to the 322\,MHz clock of the CMAC. The fast path forms the frame in a small block on the CMAC clock itself and needs neither FIFO. In either case HSB carries the control plane: address assignment by the Bootstrap Protocol (BOOTP) and register access over the User Datagram Protocol (UDP). A measurement uses one route, selected by the host software.

The fast path exists to do without the two FIFOs of HSB. An IP that runs on its own clock and streams frames of any length needs them: they carry each frame between the two clocks, and on transmit they hold it until it is complete, so that it reaches the CMAC without a gap. This workload needs neither. A request is one beat of payload behind headers that are known in advance, so the fast path, which runs on the CMAC clock, has no clock to cross and can send the frame in consecutive cycles with nothing buffered (Sec.~\ref{sec:fastpath}). The price is \emph{generality}: the fast path sends one fixed message and holds no queue. That is a deliberate trade, because this design focuses on latency.

\subsection{Remote Procedure Calls over RDMA}
A request is a CUDA-Q Realtime message~\cite{cudaqproto}: a header that carries, among other fields, a function identifier and a request identifier, followed by the arguments. The reply is a header that echoes the request identifier, followed by the result. The request identifier is the index of the call, so a late reply cannot be mistaken for the answer to a later request. The workload is the decoding loop of a memory experiment reduced to its two calls~\cite{cudaqqec}: \enq{} hands the decoder one round of syndrome bits, and \getc{} asks for the corrections it holds. Each is a separate request and reply, and a request occupies 64 bytes, one beat of the CMAC data bus.

RoCEv2 carries the InfiniBand transport in UDP datagrams~\cite{roce}. An RDMA WRITE names an address in memory that the receiver has registered, and the adapter places the payload there without involving the processor of the receiver. A WRITE with immediate also consumes a posted receive and raises a completion. A SEND delivers its payload to a posted receive. A small payload can be sent \emph{inline}, copied into the send request itself, so that the adapter does not have to fetch it from memory. Each end of an RDMA connection is a \emph{queue pair} (QP), a send queue and a receive queue. We use the \emph{unreliable connected} (UC) service, which has neither acknowledgments nor retransmission, and accept a lost frame for the sake of latency, as the NVQLink design does~\cite{nvq}.

The pattern is an established one. In HERD, clients write requests into server memory that the server polls, and the server answers with SENDs~\cite{herd}; FaRM polls ring buffers written by RDMA~\cite{farm}. The host software follows published practice for low-latency RDMA: inline payloads, unsignaled sends and a QP that is not shared between cores~\cite{kalia16}.

\subsection{FPGA Fast Path}\label{sec:fastpath}
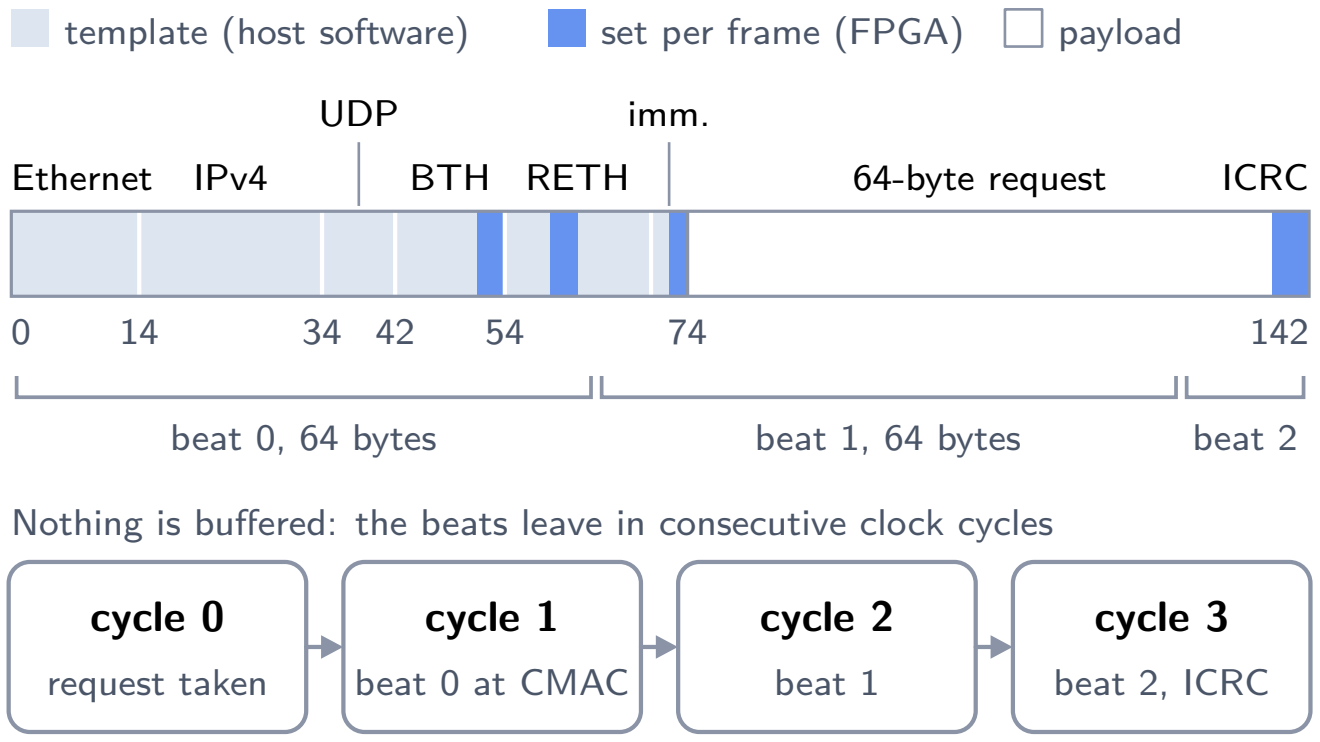
\begin{figure}[t]
\centering
\setlength{\figu}{\dimexpr\columnwidth/246\relax}%
\begin{tikzpicture}[fig, x=\figu, y=-\figu]
  \useasboundingbox (0,0) rectangle (246,140);
  \fill[figwarm] (1,2) rectangle (8,9);
  \node[anchor=base west] at (11,8.5) {template (host software)};
  \fill[figaccent!60] (102,2) rectangle (109,9);
  \node[anchor=base west] at (112,8.5) {set per frame (FPGA)};
  \draw[figline, line width=0.5pt] (188,2) rectangle (195,9);
  \node[anchor=base west] at (198,8.5) {payload};
  \node[figink, anchor=base west] at (1,36) {Ethernet};
  \node[figink] at (42.3,36) {IPv4};
  \node[figink] at (66.4,24) {UDP};
  \draw[figline, line width=0.5pt, line cap=butt] (66.4,27) -- (66.4,39);
  \node[figink] at (83.6,36) {BTH};
  \node[figink] at (107.6,36) {RETH};
  \node[figink] at (124.8,24) {imm.};
  \draw[figline, line width=0.5pt, line cap=butt] (124.8,27) -- (124.8,39);
  \node[figink] at (183.3,36) {64-byte request};
  \node[figink, anchor=base east] at (245.2,36) {ICRC};
  \fill[figwarm] (1,40) rectangle (128.28,56);
  \fill[figaccent!60] (88.72,40) rectangle (93.88,56);
  \fill[figaccent!60] (102.48,40) rectangle (107.64,56);
  \fill[figaccent!60] (124.84,40) rectangle (128.28,56);
  \fill[figaccent!60] (238.36,40) rectangle (245.24,56);
  \foreach \x in {25.08,59.48,73.24,93.88,121.4}
    \draw[white, line width=0.8pt, line cap=butt] (\x,40) -- (\x,56);
  \draw[figline, line width=0.6pt, line cap=butt] (128.28,40) -- (128.28,56);
  \draw[figline, line width=0.6pt, line join=miter] (1,40) rectangle (245.24,56);
  \node[anchor=base west] at (1,65) {0};
  \node at (25.08,65) {14};
  \node at (59.48,65) {34};
  \node at (73.24,65) {42};
  \node at (93.88,65) {54};
  \node at (128.28,65) {74};
  \node[anchor=base east] at (245.24,65) {142};
  \draw[figbr] (2,71) -- (2,75) -- (110.1,75) -- (110.1,71);
  \node at (56,85) {beat 0, 64 bytes};
  \draw[figbr] (112.1,71) -- (112.1,75) -- (220.2,75) -- (220.2,71);
  \node at (166,85) {beat 1, 64 bytes};
  \draw[figbr] (222.2,71) -- (222.2,75) -- (244.2,75) -- (244.2,71);
  \node at (233.2,85) {beat 2};
  \node[anchor=base west] at (1,101) {Nothing is buffered: the beats leave in consecutive clock cycles};
  \draw[figbox] (0.5,106) rectangle (56.5,138);
  \draw[figbox] (63.5,106) rectangle (119.5,138);
  \draw[figbox] (126.5,106) rectangle (182.5,138);
  \draw[figbox] (189.5,106) rectangle (245.5,138);
  \draw[figone] (56.5,122) -- (63.5,122);
  \draw[figone] (119.5,122) -- (126.5,122);
  \draw[figone] (182.5,122) -- (189.5,122);
  \node[fighead] at (28.5,119) {cycle 0};
  \node at (28.5,131) {request taken};
  \node[fighead] at (91.5,119) {cycle 1};
  \node at (91.5,131) {beat 0 at CMAC};
  \node[fighead] at (154.5,119) {cycle 2};
  \node at (154.5,131) {beat 1};
  \node[fighead] at (217.5,119) {cycle 3};
  \node at (217.5,131) {beat 2, ICRC};
\end{tikzpicture}
\caption{The request frame, to scale, in the form with the immediate (imm.). The host software writes the headers once as a template: Ethernet, Internet Protocol version 4 (IPv4), UDP, the base transport header (BTH) and the RDMA extended transport header (RETH). For each frame the FPGA sets only the 12 bytes shown in blue: the PSN in the BTH, the ring slot in the virtual address of the RETH and in the immediate, and the ICRC, which it derives without a pass over the frame. A frame spans three 64-byte beats of the CMAC data bus, which leave in consecutive clock cycles; cycle 0 is the cycle after the RPC client issues the request. The CMAC appends the frame check sequence. The form without the immediate omits its four bytes.}
\label{fig:frame}
\end{figure}

HSB adds the RoCEv2 headers and the invariant CRC (ICRC) to the 64-byte beat of the RPC client and emits a WRITE with immediate. On the fast path a second RPC client runs on the CMAC clock. The host software writes the headers of the request, from Ethernet to the immediate, into the FPGA once, as a template. For each frame the FPGA changes two fields: the \emph{packet sequence number} (PSN), and the ring slot, which appears in the virtual address and in the immediate (Fig.~\ref{fig:frame}). No header is computed in the FPGA. The fast path sends a request in one of two forms. With the \emph{immediate}, the request is the WRITE with immediate that HSB sends and that Fig.~\ref{fig:frame} shows, and the adapter raises a completion for it. \emph{Alone}, the WRITE carries no immediate: the frame is four bytes shorter, the ring slot appears only in the virtual address, and no completion is raised. We settle on the WRITE alone, and use the form with the immediate only for a like-for-like comparison with HSB (Sec.~\ref{sec:results}).

The ICRC would ordinarily require a pass over the frame. A CRC with a fixed initial value is, however, affine over the Galois field GF(2), the field of single bits in which addition is exclusive-or. The ICRC of a frame therefore equals that of the template with its variable fields set to zero, which the host software computes once, combined by exclusive-or with one constant for each set bit of the PSN, the ring slot and the payload. The constants depend only on the position of each bit in the frame and are fixed when the design is built. The two forms differ only in their headers, by the opcode, the length fields and the four bytes of the immediate, so the same derivation serves both, each with its own template value and its own constants for the header fields. Incremental CRC update is a known technique~\cite{braun}; we use it here so that no byte of the frame has to be buffered, and the three beats of a frame leave in consecutive clock cycles (Fig.~\ref{fig:frame}).

On receive, a tap observes the output stream of the CMAC without holding it. It compares the addressing fields of the first beat of a frame and, on a match, delivers the payload to the RPC client two cycles after that beat. It does not check the ICRC and relies on the frame check sequence of the CMAC, which covers the same bytes (Sec.~\ref{sec:limitations}). A multiplexer in front of the CMAC gives fast-path frames priority without interrupting a frame of the control plane.

Open FPGA network stacks that implement RoCE for other purposes are general engines~\cite{strom,coyote,reconic,corundum}; the fast path is not. The complete design, HSB included, occupies a small fraction of the device, and the frames of the fast path were verified bit for bit against a software reference model before the hardware was used.

\begin{table*}[t]
\caption{Round-trip times from the FPGA to the Spark and back, in \us. Each set comprises nine bursts of 200 timed calls. The median is that of the nine burst medians, with their range in parentheses. The 99th percentile (p99) is the median of the nine burst 99th percentiles. Minima, means and counts cover all 1,800 calls of a set. Every set passes through the CUDA-Q decoding session. Fast Path: Immediate is the like-for-like set, which sends a WRITE with immediate as HSB does; Fast Path: Alone is the fast path as we use it, which sends a WRITE alone. Bold marks the range quoted in the abstract.}
\label{tab:rt}
\centering\small\setlength{\tabcolsep}{2pt}
\begin{tabular*}{\linewidth}{@{\extracolsep{\fill}}lrrrrr@{}}
\toprule
FPGA path & Min. & Mean & Median (range) & p99 & Below \rt{2}\\
\midrule
HSB & 1.96 & 2.25 & 2.23 (2.19--2.28) & 2.48 & 10/1,800 (0.6\%)\\
Fast Path: Immediate & 1.72 & 2.00 & 1.99 (1.98--2.03) & 2.22 & 939/1,800 (52\%)\\
Fast Path: Alone & 1.66 & 1.91 & 1.90 (\textbf{1.87--1.95}) & 2.06 & 1,749/1,800 (\textbf{97\%})\\
\bottomrule
\end{tabular*}
\par\smallskip\raggedright\footnotesize The threshold of the last column is \rt{2.000} on HSB and 644 clock cycles, \rt{1.998}, on the fast path.
\end{table*}

\subsection{Host Software}
The relay decodes nothing itself. The FPGA fills the slots of the ring in order. The relay arms a slot by clearing the first word of the request area and writing a marker over its last bytes. A request has arrived when both have changed. The adapter places a WRITE in memory before it raises any completion, so memory is the earliest signal available to software. With a WRITE with immediate, each request also raises a completion, which the relay reads afterwards as a check. With a WRITE alone there is no completion at all, and memory is the only signal.

A handler is named by the function identifier of the request. The relay, which is written in Zig, reaches the decoding session through a C ABI: the handlers belong to a small library behind a C interface that links the CUDA-Q QEC library~\cite{cudaqqec} and holds one decoding session per configured decoder, without sockets or threads of its own. The relay calls a handler in-process, and the handler operates on the request in place.

The reply leaves as one inline SEND on the QP that received the request. Because the reply is short and its payload inline, it forms a work request of a single block, which the adapter accepts in one write to its device register~\cite{linuxrdma}. Everything that can wait does so until the reply has been posted: re-arming the slot, reading completions and re-posting receives. The connection is established without a connection manager: the host software assigns the address of the FPGA and writes the parameters of the QP, the memory key of the ring and the header template into the FPGA over the control plane.

The relay occupies the place of the dispatcher in CUDA-Q Realtime: both take a message in the CUDA-Q format, find the handler by its function identifier, call it and return the response~\cite{cudaqproto}. The dispatcher was introduced as persistent CUDA kernels on the GPU that receive, dispatch and send in separate stages~\cite{cudaqrt}, and the decoding server of CUDA-Q QEC currently dispatches from the CPU, over UDP or RDMA~\cite{cudaqserver}; that server over RDMA, with the same FPGA IP and family of adapter as our HSB path, is the closest counterpart of the relay. The dispatcher is general: its transport can be exchanged and its handlers can run on the GPU. The relay gives up that generality for a shorter path, in which one CPU thread reads the request from the memory that the adapter wrote, calls the handler in place and posts the reply on a single QP. We did not run the CUDA-Q dispatcher on the Spark, so the comparison is one of design and not of measured latency.

\section{Measurement method}\label{sec:method}
All round trips are timed by a timer in the FPGA. The timer starts in the clock cycle in which the RPC client issues a request and stops in the cycle in which it accepts the reply, so both crossings of the cable and the complete response of the Spark, hardware and host software together, lie inside the interval. Because both events are timed in the FPGA, no clock needs to be synchronized with the Spark. The resolution is one clock cycle, \ns{4} on HSB and \ns{3.1} on the fast path.

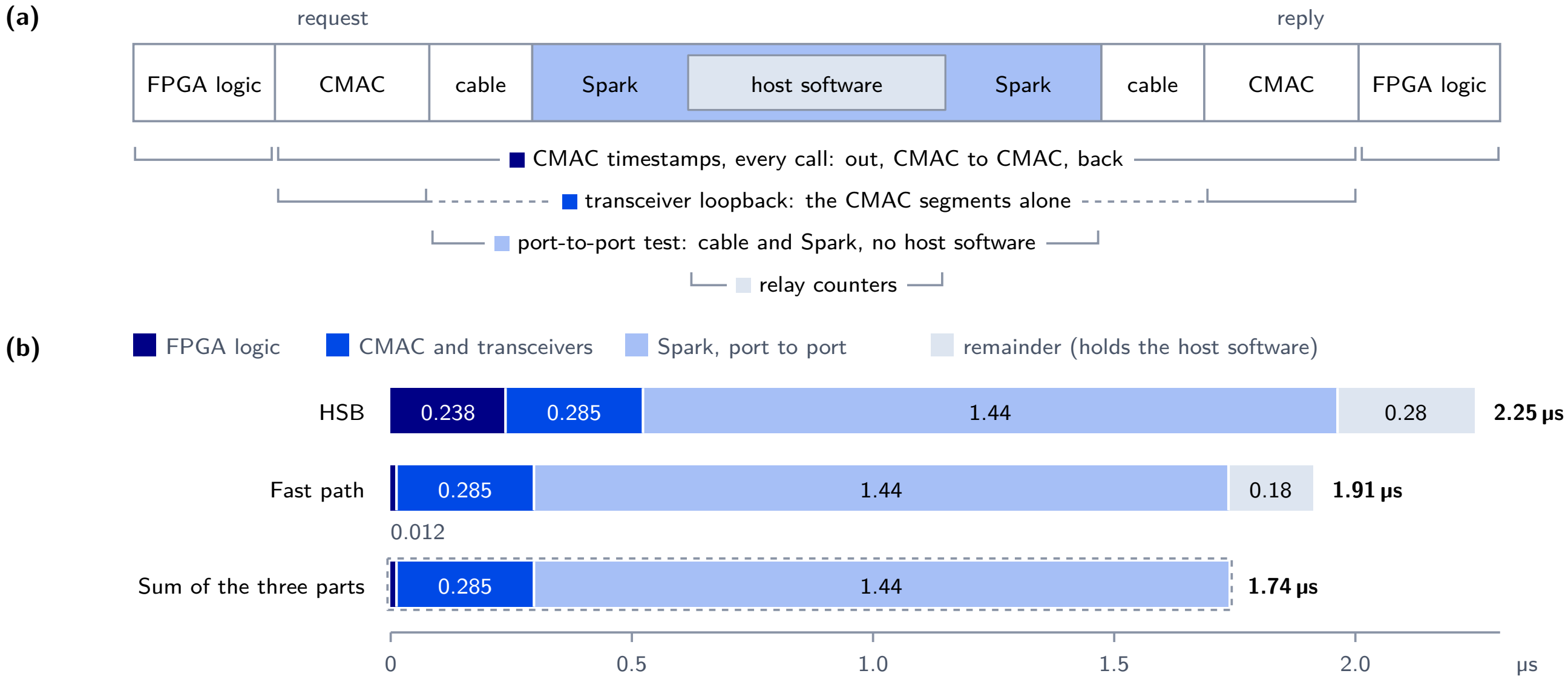
\begin{figure*}[t]
\centering
\setlength{\figu}{\dimexpr\textwidth/510\relax}%
\begin{tikzpicture}[fig, x=\figu, y=-\figu]
  \useasboundingbox (0,0) rectangle (510,214);
  \node[fighead, anchor=base west] at (0,10) {(a)};
  \node at (102,10) {request};
  \node at (403,10) {reply};
  \fill[figsky] (164,16) rectangle (341,40);
  \draw[figline, line width=0.7pt, line join=miter] (40,16) rectangle (465,40);
  \foreach \x in {84,132,164,341,373,421}
    \draw[figline, line width=0.7pt, line cap=butt] (\x,16) -- (\x,40);
  \node[figink] at (62,31) {FPGA logic};
  \node[figink] at (108,31) {CMAC};
  \node[figink] at (148,31) {cable};
  \draw[figline, line width=0.6pt, fill=figwarm] (212.5,19.5) rectangle (292.5,36.5);
  \node[figink] at (188.25,31) {Spark};
  \node[figink] at (252.5,31) {host software};
  \node[figink] at (316.75,31) {Spark};
  \node[figink] at (357,31) {cable};
  \node[figink] at (397,31) {CMAC};
  \node[figink] at (443,31) {FPGA logic};
  \draw[figbr] (40.5,48) -- (40.5,52) -- (83,52) -- (83,48);
  \draw[figbr] (85,48) -- (85,52) -- (420,52) -- (420,48);
  \draw[figbr] (422,48) -- (422,52) -- (464.5,52) -- (464.5,48);
  \node[figinst] at (252.5,52) {{\color{figdark}\rule[-0.4pt]{4.6pt}{4.6pt}}\hspace{2.5pt}CMAC timestamps, every call: out, CMAC to CMAC, back};
  \draw[figbr] (85,61) -- (85,65) -- (131,65) -- (131,61);
  \draw[figbr] (374,61) -- (374,65) -- (420,65) -- (420,61);
  \draw[figbr, dash pattern=on 2pt off 2pt] (131,65) -- (374,65);
  \node[figinst] at (252.5,65) {{\color{figmid}\rule[-0.4pt]{4.6pt}{4.6pt}}\hspace{2.5pt}transceiver loopback: the CMAC segments alone};
  \draw[figbr] (133,74) -- (133,78) -- (340,78) -- (340,74);
  \node[figinst] at (236.5,78) {{\color{figsky}\rule[-0.4pt]{4.6pt}{4.6pt}}\hspace{2.5pt}port-to-port test: cable and Spark, no host software};
  \draw[figbr] (213.5,87) -- (213.5,91) -- (291.5,91) -- (291.5,87);
  \node[figinst] at (252.5,91) {{\color{figwarm}\rule[-0.4pt]{4.6pt}{4.6pt}}\hspace{2.5pt}relay counters};
  \node[fighead, anchor=base west] at (0,113) {(b)};
  \fill[figdark] (40,106) rectangle (47,113);
  \node[anchor=base west] at (50,112.5) {FPGA logic};
  \fill[figmid] (100,106) rectangle (107,113);
  \node[anchor=base west] at (110,112.5) {CMAC and transceivers};
  \fill[figsky] (193,106) rectangle (200,113);
  \node[anchor=base west] at (203,112.5) {Spark, port to port};
  \fill[figwarm] (288,106) rectangle (295,113);
  \node[anchor=base west] at (298,112.5) {remainder (holds the host software)};
  \node[figink, anchor=base east] at (112,133) {HSB};
  \fill[figdark] (120.00,123) rectangle (155.70,137);
  \node[text=white] at (137.85,133) {0.238};
  \fill[figmid] (155.70,123) rectangle (198.45,137);
  \node[text=white] at (177.07,133) {0.285};
  \draw[white, line width=0.8pt, line cap=butt] (155.70,123) -- (155.70,137);
  \fill[figsky] (198.45,123) rectangle (414.45,137);
  \node[figink] at (306.45,133) {1.44};
  \draw[white, line width=0.8pt, line cap=butt] (198.45,123) -- (198.45,137);
  \fill[figwarm] (414.45,123) rectangle (457.00,137);
  \node[figink] at (435.73,133) {0.28};
  \draw[white, line width=0.8pt, line cap=butt] (414.45,123) -- (414.45,137);
  \node[figink, anchor=base west, font=\sffamily\scriptsize\bfseries] at (463.00,133) {2.25\,\textmu s};
  \node[figink, anchor=base east] at (112,157) {Fast path};
  \fill[figdark] (120.00,147) rectangle (121.80,161);
  \node[anchor=base west] at (120.00,170) {0.012};
  \fill[figmid] (121.80,147) rectangle (164.55,161);
  \node[text=white] at (143.17,157) {0.285};
  \draw[white, line width=0.8pt, line cap=butt] (121.80,147) -- (121.80,161);
  \fill[figsky] (164.55,147) rectangle (380.55,161);
  \node[figink] at (272.55,157) {1.44};
  \draw[white, line width=0.8pt, line cap=butt] (164.55,147) -- (164.55,161);
  \fill[figwarm] (380.55,147) rectangle (406.80,161);
  \node[figink] at (393.68,157) {0.18};
  \draw[white, line width=0.8pt, line cap=butt] (380.55,147) -- (380.55,161);
  \node[figink, anchor=base west, font=\sffamily\scriptsize\bfseries] at (412.80,157) {1.91\,\textmu s};
  \node[figink, anchor=base east] at (112,187) {Sum of the three parts};
  \fill[figdark] (120.00,177) rectangle (121.80,191);
  \fill[figmid] (121.80,177) rectangle (164.55,191);
  \node[text=white] at (143.17,187) {0.285};
  \draw[white, line width=0.8pt, line cap=butt] (121.80,177) -- (121.80,191);
  \fill[figsky] (164.55,177) rectangle (380.55,191);
  \node[figink] at (272.55,187) {1.44};
  \draw[white, line width=0.8pt, line cap=butt] (164.55,177) -- (164.55,191);
  \draw[figline, line width=0.7pt, line cap=butt, dash pattern=on 2pt off 2pt] (119.00,176) rectangle (381.55,192);
  \node[figink, anchor=base west, font=\sffamily\scriptsize\bfseries] at (386.55,187) {1.74\,\textmu s};
  \draw[figline, line width=0.7pt, line cap=butt] (120,199) -- (465,199);
  \draw[figline, line width=0.7pt, line cap=butt] (120,199) -- (120,202);
  \node at (120,211) {0};
  \draw[figline, line width=0.7pt, line cap=butt] (195,199) -- (195,202);
  \node at (195,211) {0.5};
  \draw[figline, line width=0.7pt, line cap=butt] (270,199) -- (270,202);
  \node at (270,211) {1.0};
  \draw[figline, line width=0.7pt, line cap=butt] (345,199) -- (345,202);
  \node at (345,211) {1.5};
  \draw[figline, line width=0.7pt, line cap=butt] (420,199) -- (420,202);
  \node at (420,211) {2.0};
  \node[anchor=base west] at (470,211) {\textmu s};
\end{tikzpicture}
\caption{Latency budget of a round trip. (a) One round trip along its path, not to scale, with the span that each instrument covers. The square beside each of the first three instruments marks the part of (b) that it yields; the relay counters time the host software, which lies inside the last part. The FPGA timer spans the whole strip. The CMAC timestamps act on every call; the other three are separate measurements. (b) The mean round trip through the decoding session, to scale and in \us, for HSB and the fast path (first and last rows of Table~\ref{tab:rt}), divided into the three parts that are measured independently and the remainder, which holds the host software. The two bars are different sets, and of the difference between them only the FPGA logic is attributed to the path (Sec.~\ref{sec:results}). The Spark part is its port-to-port turnaround, measured with none of the host software. The last bar is the sum of the three parts, which is not measured as one. Values are rounded independently.}
\label{fig:budget}
\end{figure*}

Four instruments divide this interval, and Fig.~\ref{fig:budget}(a) shows the span of each. First, two timestamps at the boundary of the CMAC split every round trip into three spans: FPGA logic outbound, CMAC to CMAC around the cable and the Spark, and FPGA logic inbound. Second, near-end loopback of the transceivers returns a request from the transmit side of the CMAC to its receive side, which isolates the CMAC and transceivers. Third, a port-to-port test runs a standard RDMA latency benchmark~\cite{linuxrdma} between the two ports of the adapter of the Spark, with the UC service, 64-byte inline messages and none of the host software. The benchmark reports half of its own round trip, which we take from its documentation to be one turnaround of the Spark, through its adapter and processor, and one crossing of the cable. We call this figure the port-to-port turnaround of the Spark. Fourth, when enabled, counters in the relay time its stages from seeing a request to posting the reply.

Requests alternate between the two calls, 1\,ms apart with one request outstanding at a time, and the FPGA checks every reply against the result expected from the identity decoder. A burst consists of a few untimed warm-up calls and 200 timed calls. A measurement set comprises three runs of three bursts, with the relay restarted between runs, and therefore 1,800 timed calls. For each burst the FPGA reports the minimum, the mean and the number of calls at or above a threshold of \rt{2.00}. It also stores the time of every call, from which the host software derives the median and the 99th percentile, which with 200 samples is the second-largest value. Table~\ref{tab:rt} summarizes a set by the medians of these two over its nine bursts.

\section{Results}\label{sec:results}
\subsection{Round Trip}
Table~\ref{tab:rt} gives the measured round trips. The first two rows of the table are a like-for-like comparison. They use the same decoding session, the same 64-byte request, the same RDMA operation, WRITE with immediate, and the same cable. Only the FPGA path differs. The CMAC timestamps put the contribution of the FPGA logic at \ns{238} on HSB and at \ns{12} on the fast path. The median round trip falls with it, from 2.23 to \rt{1.99}, while the CMAC-to-CMAC span is nearly unchanged. The \ns{238} is the cost of HSB as integrated here, with its clock crossing and its store-and-forward FIFO, not of the vendor IP alone. The measurement does not separate the two.

That shift moves a large fraction of the replies across \rt{2}: about half of the calls in the like-for-like set complete below it, against almost none on HSB.

The fast path itself sends a WRITE alone, which removes one completion per request from the work of the adapter and the relay. On the fast path (last row of Table~\ref{tab:rt}) the burst medians lie between 1.87 and \rt{1.95}, and 97\% of the calls complete below \rt{2}. Repeated sets of the same configuration differed by up to \rt{0.09} in the median, so we do not credit the difference between the last two rows to the RDMA operation.

\subsection{Latency Budget}\label{sec:budget}
Three parts that contain none of the host software account for most of the round trip, as Table~\ref{tab:budget} and Fig.~\ref{fig:budget}(b) show.

The FPGA logic contributes \ns{12}, four clock cycles, two in each direction, and the same value in every fast-path burst. In loopback the CMAC-to-CMAC span is \ns{285} on average: the transmit side of the CMAC, the transceivers and the receive side together. The CMAC is a hard block, fixed in the silicon of the FPGA, and we know of no setting that reduces its latency.

The port-to-port turnaround of the Spark, one pass through its adapter and processor with one crossing of the cable, takes \rt{1.44}. It belongs to the Spark and not to the host software, none of which runs in the port-to-port test, and it is the largest part by a factor of five.
\begin{table}[b]
\caption{Latency budget of a fast-path round trip through the decoding session, in \us. The remainder holds the host software. Entries are rounded independently.}
\label{tab:budget}
\centering\small\setlength{\tabcolsep}{3pt}
\begin{tabular*}{\linewidth}{@{\extracolsep{\fill}}lrl@{}}
\toprule
Part & Mean & Measured by\\
\midrule
FPGA logic & 0.012 & CMAC timestamps\\
CMAC and transceivers & 0.285 & loopback\\
Spark, port to port & 1.44 & port-to-port test\\
\midrule
Sum of the three parts & 1.74 & \\
Measured round trip & 1.91 & FPGA timer\\
Remainder & 0.18 & \\
\bottomrule
\end{tabular*}
\end{table}

\begin{table*}[t]
\caption{Small-message network round trips from an FPGA to a CPU or GPU and back, as reported in quantum-control papers. For this work the range of the nine burst medians of the fast path (last row of Table~\ref{tab:rt}) is given. The other rows give the medians measured in the cited papers, and the last column locates each value in its source.}
\label{tab:lit}
\centering\small\setlength{\tabcolsep}{4pt}
\begin{tabular*}{\linewidth}{@{\extracolsep{\fill}}lp{.56\linewidth}ll@{}}
\toprule
System & Responder and benchmark & Median (\us) & Location in source\\
\midrule
This work & DGX Spark CPU, CUDA-Q decoding session, 64-byte request, 100\,GbE & 1.87--1.95 & Table~\ref{tab:rt} here\\
Backline~\cite{backline} & Threadripper PRO CPU, 16-byte echo, software-scheduled rounds & 2.305 & Fig.~6\\
Backline~\cite{backline} & Instinct MI210 GPU, same echo and scheduling & 4.500 & Fig.~6\\
Backline~\cite{backline} & Threadripper PRO CPU, FPGA-scheduled back-to-back echo & 2.470 & App.\ A, Table~X\\
QDPU~\cite{qdpu} & DGX Spark CPU, 128-byte forwarding at 10\,GbE & 2.74 & Table~42\\
QDPU~\cite{qdpu} & DGX Spark GPU, same 128-byte forwarding test & 6.40 & Table~42\\
THQLink~\cite{thq} & High-performance computing host, round trip over TH-Express & 2.870 & Fig.~4\\
NVQLink~\cite{nvq} & GPU echo kernel, 32-byte payload & 3.839 & Fig.~5\\
\bottomrule
\end{tabular*}
\par\smallskip\raggedright\footnotesize Wang et al.\ also quote a later NVIDIA median of \rt{2.92} for GPU forwarding (their Appendix~A; payload size unspecified)~\cite{wang}.
\end{table*}

The three parts sum to \rt{1.74}, against a measured mean of \rt{1.91} on the fast path. The remaining \rt{0.18}, less than a tenth of the round trip, holds everything else, the host software included. The relay counters measure the host software directly: from seeing a request to posting the reply, the relay and the decoding session together take 0.13 to \rt{0.22} per call. These times do not simply add to the remainder, because the send and poll of the benchmark are already inside the \rt{1.44}. About 85\% of the round trip is thus spent beyond the CMAC of the FPGA, and most of that in the Spark, below the host software.

\section{Comparison with prior work}\label{sec:comparison}
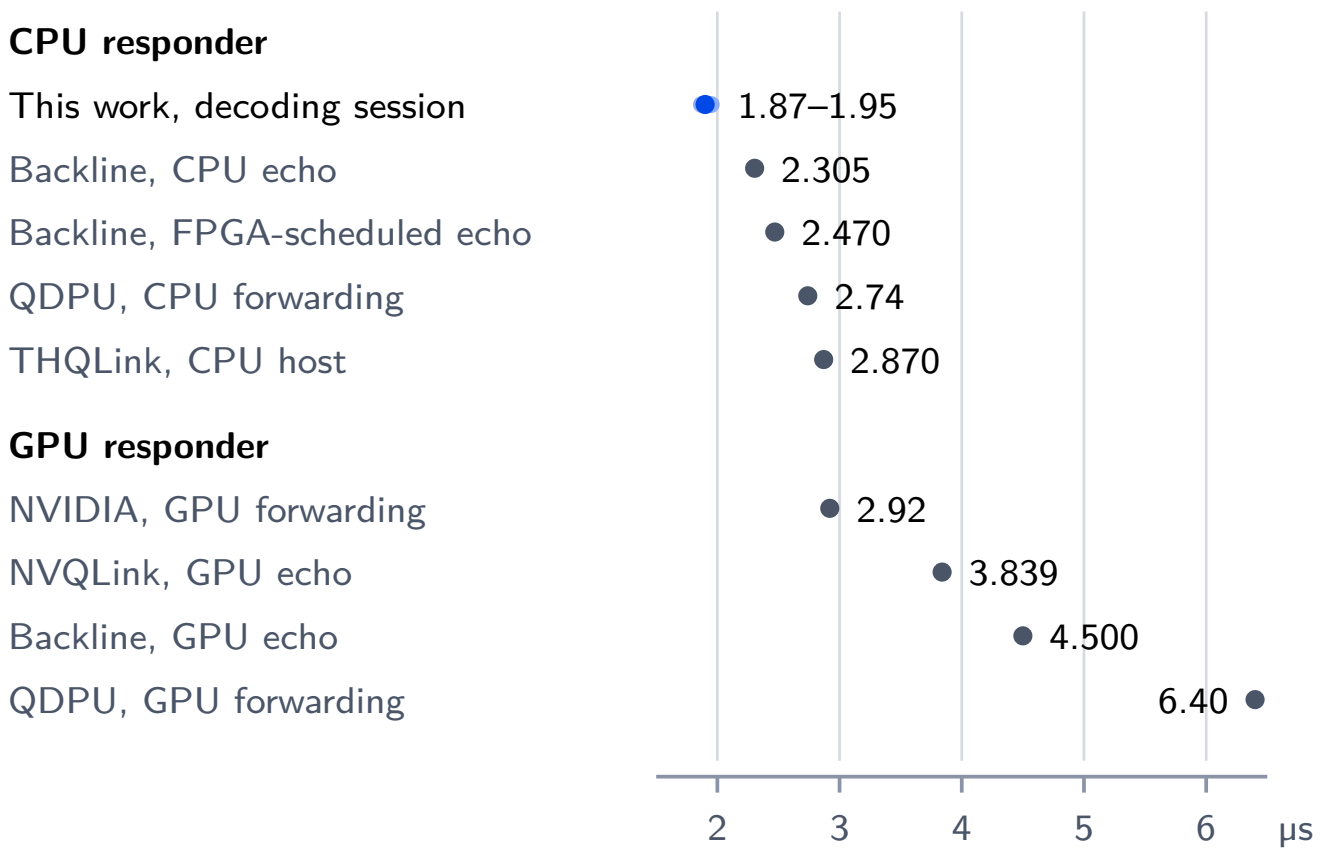
\begin{figure}[t]
\centering
\setlength{\figu}{\dimexpr\columnwidth/246\relax}%
\begin{tikzpicture}[fig, x=\figu, y=-\figu]
  \useasboundingbox (0,0) rectangle (246,161);
  \draw[figline!35, line width=0.5pt, line cap=butt] (133.5,2) -- (133.5,143);
  \draw[figline!35, line width=0.5pt, line cap=butt] (156.5,2) -- (156.5,143);
  \draw[figline!35, line width=0.5pt, line cap=butt] (179.5,2) -- (179.5,143);
  \draw[figline!35, line width=0.5pt, line cap=butt] (202.5,2) -- (202.5,143);
  \draw[figline!35, line width=0.5pt, line cap=butt] (225.5,2) -- (225.5,143);
  \node[figink, anchor=base west, font=\sffamily\scriptsize\bfseries] at (0,10) {CPU responder};
  \node[figink, anchor=base west] at (0,22) {This work, decoding session};
  \draw[figaccent!45, line width=3.2pt, line cap=round] (130.51,19.5) -- (132.35,19.5);
  \fill[figaccent] (131.20,19.5) circle[radius=1.8pt];
  \node[figink, anchor=base west] at (137.35,22) {1.87--1.95};
  \node[anchor=base west] at (0,34) {Backline, CPU echo};
  \fill[figquiet] (140.52,31.5) circle[radius=1.8pt];
  \node[figink, anchor=base west] at (145.52,34) {2.305};
  \node[anchor=base west] at (0,46) {Backline, FPGA-scheduled echo};
  \fill[figquiet] (144.31,43.5) circle[radius=1.8pt];
  \node[figink, anchor=base west] at (149.31,46) {2.470};
  \node[anchor=base west] at (0,58) {QDPU, CPU forwarding};
  \fill[figquiet] (150.52,55.5) circle[radius=1.8pt];
  \node[figink, anchor=base west] at (155.52,58) {2.74};
  \node[anchor=base west] at (0,70) {THQLink, CPU host};
  \fill[figquiet] (153.51,67.5) circle[radius=1.8pt];
  \node[figink, anchor=base west] at (158.51,70) {2.870};
  \node[figink, anchor=base west, font=\sffamily\scriptsize\bfseries] at (0,86) {GPU responder};
  \node[anchor=base west] at (0,98) {NVIDIA, GPU forwarding};
  \fill[figquiet] (154.66,95.5) circle[radius=1.8pt];
  \node[figink, anchor=base west] at (159.66,98) {2.92};
  \node[anchor=base west] at (0,110) {NVQLink, GPU echo};
  \fill[figquiet] (175.80,107.5) circle[radius=1.8pt];
  \node[figink, anchor=base west] at (180.80,110) {3.839};
  \node[anchor=base west] at (0,122) {Backline, GPU echo};
  \fill[figquiet] (191.00,119.5) circle[radius=1.8pt];
  \node[figink, anchor=base west] at (196.00,122) {4.500};
  \node[anchor=base west] at (0,134) {QDPU, GPU forwarding};
  \fill[figquiet] (234.70,131.5) circle[radius=1.8pt];
  \node[figink, anchor=base east] at (229.70,134) {6.40};
  \draw[figline, line width=0.7pt, line cap=butt] (122.0,146) -- (237.0,146);
  \draw[figline, line width=0.7pt, line cap=butt] (133.5,146) -- (133.5,149);
  \node at (133.5,158) {2};
  \draw[figline, line width=0.7pt, line cap=butt] (156.5,146) -- (156.5,149);
  \node at (156.5,158) {3};
  \draw[figline, line width=0.7pt, line cap=butt] (179.5,146) -- (179.5,149);
  \node at (179.5,158) {4};
  \draw[figline, line width=0.7pt, line cap=butt] (202.5,146) -- (202.5,149);
  \node at (202.5,158) {5};
  \draw[figline, line width=0.7pt, line cap=butt] (225.5,146) -- (225.5,149);
  \node at (225.5,158) {6};
  \node[anchor=base west] at (239.0,158) {\textmu s};
\end{tikzpicture}
\caption{The round trips of Table~\ref{tab:lit}, by responder. For this work the bar spans the nine burst medians of the fast path and the mark is their median. The other values are the medians given by each source. NVIDIA denotes the later value quoted by Wang et al.~\cite{wang}. Conditions differ between experiments.}
\label{fig:lit}
\end{figure}

Table~\ref{tab:lit} and Fig.~\ref{fig:lit} compare small-message network round trips from an FPGA to a CPU or GPU and back, as reported in quantum-control papers. The comparison is limited to the papers listed, and we make no claim about results reported elsewhere. Backline reports a CPU echo median of \rt{2.305}, the lowest median for an external CPU among them~\cite{backline}. Every burst median of our fast-path sets through the decoding session is lower. The other entries, for CPU and for GPU responders, lie higher~\cite{backline,qdpu,thq,nvq,wang}.

These values are not directly comparable. The responders, the handlers, the payloads, the pacing and the statistics differ between experiments, and the results with a GPU responder have no counterpart here. The controlled comparison is the one within Table~\ref{tab:rt}, between HSB and the fast path, which changes only the FPGA path.

\section{Limitations}\label{sec:limitations}
The results are bench measurements on one set of hardware and hold under the following conditions.

\emph{Pacing.} Requests were issued once per millisecond, one at a time, which is the pattern the fast path is built for: it holds no queue. Back-to-back operation and overlapping requests at syndrome rates remain to be measured.

\emph{Workload.} A decoding step carries three syndrome bits. As stated at the outset, the session runs an identity decoder on a CPU thread, with no path into the GPU, so the times are those of the link and of the session interface. The two calls are separate round trips, one outstanding at a time, so a decision that waits for both costs two of them.

\emph{Hardware.} One board, one cable and one Spark with its adapter were used, on a direct link without a switch or forward error correction; the type and length of the cable were not recorded. A switch would add to the round trip; Ref.~\cite{thq} reports \rt{0.13} per hop on its own interconnect. Sets were measured in sequence, and Sec.~\ref{sec:results} gives the variation between them.

\emph{Integrity.} For replies the fast path relies on the frame check sequence in place of the ICRC, which is sufficient on a direct link. A switched network would call for the ICRC check as well.

\makeatletter\@secpenalty=10000\makeatother
\section{Conclusion and outlook}\label{sec:conclusion}
In summary, we have realized a customized network link between an FPGA and a GPU server (Spark), in which a round trip from a control FPGA through a decoding session on a commodity machine and back is measured to have a sub-\rt{2} median latency: burst medians of 1.87 to \rt{1.95} on the fast path, and a median of \rt{1.99} against \rt{2.23} in the like-for-like comparison with HSB. About 85\% of that time is spent beyond the CMAC, most of it in the Spark itself.

The protocol work of the FPGA can be reduced to \ns{12} by sending a header written by the host software and deriving the ICRC from the linearity of the CRC. What remains is \ns{285} in the CMAC and its transceivers, which we did not attempt to modify, and \rt{1.44} for the port-to-port turnaround of the Spark, which was measured using a standard benchmarking method without the involvement of the host software. The method needs only the two endpoints, and its four instruments are available on any such link whose adapter has a second port.

In a future experiment, one could consider sharing a GPU decoder server via a switch among multiple FPGA controllers at their respective syndrome rates, and measure the interval from the last required measurement to the conditional pulse. The fraction of gates that receive a correct decision in time will show how a sub-\rt{2} median round trip translates into a scalable feedback system.

\end{document}